\documentclass[aps,prl,twocolumn,showpacs,showkeys,groupedaddress]{revtex4-1}

\usepackage{amsmath}
\usepackage{graphicx}
\usepackage{color}
\usepackage{amssymb}

\def\TeV{\ifmmode {\mathrm{Te\kern -0.1em V}}\else
                   \textrm{Te\kern -0.1em V}\fi}%
\def\GeV{\ifmmode {\mathrm{Ge\kern -0.1em V}}\else
                   \textrm{Ge\kern -0.1em V}\fi}%
\def\MeV{\ifmmode {\mathrm{Me\kern -0.1em V}}\else
                   \textrm{Me\kern -0.1em V}\fi}%
\def\keV{\ifmmode {\mathrm{ke\kern -0.1em V}}\else
                   \textrm{ke\kern -0.1em V}\fi}%
\def\eV{\ifmmode  {\mathrm{e\kern -0.1em V}}\else
                   \textrm{e\kern -0.1em V}\fi}%
\let\tev=\TeV
\let\gev=\GeV
\let\mev=\MeV

\def\ifb{\mbox{fb$^{-1}$}}

\def\MSbar{\mbox{$\overline{\mathrm{MS}}$\,}}

\begin{document}

\preprint{IFIC-2021-, UWThPh 2021-20}

\title{$m_b$ at $m_H$: the running bottom quark mass and the Higgs boson}



\author{Javier Aparisi}
\affiliation{Instituto de F\'isica Corpuscular, CSIC-University of Valencia, Valencia, Spain}

\author{Juan Fuster}
\affiliation{Instituto de F\'isica Corpuscular, CSIC-University of Valencia, Valencia, Spain}

\author{Andr\'e Hoang}
\affiliation{University of Vienna, Austria}

\author{Adri\'an Irles}
\affiliation{Instituto de F\'isica Corpuscular, CSIC-University of Valencia, Valencia, Spain}

\author{Christopher Lepenik}
\affiliation{University of Vienna, Austria}

\author{Germ\'an Rodrigo}
\affiliation{Instituto de F\'isica Corpuscular, CSIC-University of Valencia, Valencia, Spain}

\author{Michael Spira}
\affiliation{PSI Villigen, Switzerland}

\author{Seidai Tairafune}
\affiliation{Tohoku University, Sendai, Japan}


\author{Marcel Vos}
\email[]{marcel.vos@ific.uv.es}
\affiliation{Instituto de F\'isica Corpuscular, CSIC-University of Valencia, Valencia, Spain}

\author{Hitoshi Yamamoto}
\altaffiliation{On leave from Tohoku University, Sendai, Japan}
\affiliation{Instituto de F\'isica Corpuscular, CSIC-University of Valencia, Valencia, Spain}

\author{Ryo Yonamine}
\affiliation{Tohoku University, Sendai, Japan}


\date{\today}

\begin{abstract}
  We present a new measurement of the bottom quark mass in the $\bar{MS}$ scheme at the renormalization scale of the Higgs boson mass from measurements of Higgs boson decay rates at the LHC: $m_b(m_H)= 2.60^{+0.36}_{-0.31}$~\gev. The measurement has a negligible theory uncertainty and excellent prospects to improve at the HL-LHC and a future Higgs factory. Confronting this result, and $m_b(m_b)$ from low-energy measurements and $m_b(m_Z)$ from $Z-$pole data, with the prediction of the scale evolution of the renormalization group equations, we find strong evidence for the "running" of the bottom quark mass. 
\end{abstract}



\maketitle

\section{Introduction}

Quark masses are renormalized and scheme-dependent parameters of the Standard Model (SM) Lagrangian. Their values must be determined experimentally, through the comparison of measurements of physical observables sensitive to the mass to SM predictions in Quantum Chromodynamics (QCD) beyond leading order accuracy. In the most popular renormalization scheme, the modified minimal subtraction scheme or $\overline{\rm MS}$ scheme, the strong coupling $\alpha_s(\mu)$ and the quark masses $m_q(\mu)$ are "running constants" that depend on the dimensionful renormalization scale $\mu$. 

QCD yields a precise prescriptions for the scale evolution: given a value for a quark mass at a reference scale, its value can be determined at any other scale using the renormalization group equation (RGE). Determinations of the RGE for the the running quark masses have by now reached the 5-loop (${\cal O}(\alpha_s^5)$) level~\cite{Vermaseren:1997fq,Chetyrkin:1997dh,Baikov:2014qja} and software packages such as RunDec~\cite{Herren:2017osy} and REvolver~\cite{Hoang:2021fhn} provide access to state-of-the-art renormalization evolution and scheme conversions.

Performing several measurements at different energy scales, the renormalization scale-dependence of the strong coupling and the quark masses in the $\overline{\rm MS}$ scheme can be tested experimentally. For each measurement one can identify a typical scale of the physical process, where high-order logarithmic corrections related to renormalization group invariance are resummed and predictions yield nicely behaved perturbative series. A large number of determinations over a broad range of energies characterizes the evolution of the strong coupling $\alpha_s(\mu)$~\cite{Zyla:2020zbs}. Experiments have also found evidence for the ``running'' of \MSbar{} quark masses for the charm quark at HERA~\cite{Gizhko:2017fiu} and have studied the scale evolution of the top quark at the LHC~\cite{Sirunyan:2019jyn}. 

The most precise measurements of the mass of the bottom quark are performed at relatively low energy. The "world average" is given by the Particle Data Group, as follows:
 \begin{equation}
  m_b(m_b) = 4.18^{+0.03}_{-0.02}~\gev,
\label{eq:mbmb}
\end{equation}
where the reference value of the bottom mass is quoted in the \MSbar\ scheme, at a scale given by the mass itself~\footnote{Throughout this paper we define the $\overline{\rm MS}$ bottom quark mass in the $n_f=5$ flavor scheme.}. Measurements at the scale of the $Z$-boson mass have been performed by the LEP experiments and using SLD data~\cite{Abreu:1997ey,Rodrigo:1997gy,Abe:1998kr,Brandenburg:1999nb,Barate:2000ab,Abbiendi:2001tw,Abdallah:2005cv,Abdallah:2008ac}. We use the following average~\footnote{Details of the averaging method are given in the supplementary material.} of the most precise measurements for $m_b(m_Z)$:
\begin{equation}
    m_b(m_Z) = 2.82 \pm 0.28~\GeV.
    \label{eq:mbmz}
\end{equation}

\section{Bottom quark mass from Higgs decay}

The discovery of the Higgs boson~\cite{Aad:2012tfa,Chatrchyan:2012ufa} and the observation of Higgs boson decay to bottom quark pairs~\cite{Aaboud:2018zhk,Sirunyan:2018kst} at the LHC~\cite{Bruning:2004ej,Evans:2008zzb} enables an entirely new measurement of the bottom quark mass. 
The ATLAS~\cite{Aad:2008zzm} and CMS~\cite{Chatrchyan:2008aa} experiments have characterized the product of Higgs boson production and decay rates in many combinations of production processes and Higgs decay channels. The measurements of the $H\rightarrow b\bar{b}$ branching ratio, combining the $VH$, $t\bar{t}H$ associated production modes and the vector-boson-fusion mode, have achieved a precision of approximately 40\% in run 1~\cite{Khachatryan:2016vau} and 20\% in run 2~\cite{Sirunyan:2018koj,Aad:2019mbh,CMS-PAS-HIG-19-005,ATLAS-CONF-2020-027}. 

We focus on the ratio  $BR (H \rightarrow b\bar{b})/BR(H\rightarrow ZZ)$ of the branching ratios to bottom quarks and to $Z$ bosons. The LHC experiments present this result relative to the SM prediction of $BR(H\rightarrow b\bar{b})/BR(H \rightarrow ZZ) =$ 22.0 $\pm$ 0.5. 
The bottom quark mass measurement presented in this letter is based on a preliminary result by ATLAS~\cite{ATLAS-CONF-2020-027}, using 139~\ifb{} at $\sqrt{s}=$ 13~\tev{},
\begin{equation}
\label{eq:mubbmuzz_atlas}
 \mu^{bb}/\mu^{ZZ} =  0.87^{+0.22}_{-0.17} (stat.)  ^{+0.18}_{-0.12} (syst.) = 0.87^{+0.28}_{-0.21},
\end{equation}
and a result by CMS~\cite{Sirunyan:2018koj} based on 35~\ifb{}: 
\begin{equation}
\label{eq:mubbmuzz_cms}
 \mu^{bb}/\mu^{ZZ} =  0.84^{+0.27}_{-0.21} (stat.)  ^{+0.26}_{-0.17} (stat.) = 0.84^{+0.37}_{-0.27}.
\end{equation}

\textbf{Dependence of Higgs boson decay rates on $m_b$.} 
In the limit $m_b << m_H$, the partial decay width can be written in the form:
\begin{eqnarray}
\Gamma [H\to b\bar b] & = & \frac{3 G_{F} m_H } {4\sqrt{2}\pi}
\ m_b(\mu)^2\ \left(1+\delta_{\rm ew} \right) \nonumber \\
& \times & [1 + \delta_{\rm QCD}+\delta_t + \delta_{\rm mix}]
\label{eq:h2bb}
\end{eqnarray}
where $G_F$ denotes the Fermi constant, $\delta_{\rm QCD}$ the QCD corrections related to the scalar correlator, $\delta_t$ the QCD corrections due to the interference with $H\to gg$ diagrams that start to contribute at NNLO, $\delta_{\rm ew}$ the electroweak (EW) corrections and finally $\delta_{\rm mix}$ the mixed QCD-EW corrections. The decay width has a quadratic dependence on the bottom quark mass and can be precisely predicted. In particular, the QCD corrections $\delta_{\rm QCD}$ are known up to N$^4$LO \cite{Braaten:1980yq, Sakai:1980fa, Inami:1980qp, Drees:1990dq, Drees:1989du,Gorishnii:1983cu, Gorishnii:1990zu, Gorishnii:1991zr, Kataev:1993be, Surguladze:1994gc, Chetyrkin:1996sr, Melnikov:1995yp, Baikov:2005rw, Herzog:2017dtz}, the interference term $\delta_t$ at NNLO \cite{Chetyrkin:1995pd, Larin:1995sq, Primo:2018zby}, the EW corrections $\delta_{\rm ew}$ at NLO \cite{Fleischer:1980ub, Bardin:1990zj, Dabelstein:1991ky, Kniehl:1991ze} and finally the mixed corrections $\delta_{\rm mix}$ at two-loop order \cite{Kataev:1997cq, Kniehl:1994ju, Kwiatkowski:1994cu, Chetyrkin:1996wr, Mihaila:2015lwa, Chaubey:2019lum}.

The Higgs boson mass is the characteristic dynamical scale for the decay rate into bottom quarks. A measurement of the $H\rightarrow b\bar{b}$ partial width therefore naturally provides a measurement of the bottom quark mass at the renormalization scale of the Higgs boson mass~\cite{Braaten:1980yq,Sakai:1980fa}. This point can be illustrated by considering the convergence of the perturbative series.
When the renormalization scale $\mu=m_H$ is adopted for the strong coupling and the bottom quark mass (we use $m_b(m_b)=4.18$~GeV and $\alpha_s(m_Z)=0.1179$ as input and obtain $m_b(m_H)=2.790$~GeV, $\alpha_s(m_H)=0.1125$ and $\alpha_s(m_b(m_b))=0.2245$ using 5-loop RGEs for 5 active flavors), the leading QCD series for the $H\rightarrow b\bar{b}$ partial width (in the expansion in $m_b^2/m_H^2$) takes the following form:
\begin{equation}
\label{Gammamumh}
1 + \delta_{\rm QCD} = 1 + 0.2030 + 0.0374 + 0.0019 - 0.0014.
\end{equation}
The successive loop corrections, listed as separate terms, show excellent convergence. 
In contrast, using $\mu=m_b$, the leading perturbation series adopts the form
\begin{equation}
\label{Gammamumb}
1 + \delta_{\rm QCD} = 1 - 0.5665 + 0.0586 + 0.1475  - 0.1274, 
\end{equation}
which shows very poor convergence behavior and has large perturbative uncertainties due to powers of the large logarithm $\ln(m_H/m_b)$. These large logarithmic terms are resummed to all orders in Eq.~(\ref{Gammamumh}), which explains its much better behavior. This property supports the idea that the $H \rightarrow b\bar{b}$ partial width provides a measurement of the bottom quark mass at the renormalization scale $m_H$. A more detailed discussion is found in the supplementary material provided with this Letter.

\textbf{Numerical predictions.} The dependence of the $H\rightarrow b\bar{b}$ partial width on the bottom quark mass is obtained with HDECAY~\cite{Djouadi:2018xqq,Djouadi:1997yw}. The calculation of the $H\rightarrow b\bar{b}$ decay accounts for N$^4$LO corrections in QCD and includes NLO EW corrections. The full bottom quark mass effects are taken into account up to NLO and logarithmic ones up to NNLO. Version 6.61 of the code is used, where the bottom quark \MSbar\ mass can be supplied at the scale $\mu=m_H$ of the decay process. We use $m_H$ = 125.1~\gev{} and $\alpha_s(m_Z)=$ 0.1179 (PDG world average) throughout this letter. For a bottom quark mass $m_b(m_H) =$ 2.79~\GeV{} (corresponding to $m_b(m_b)=$ 4.18~\gev) HDECAY predicts a partial width of 2.363~\MeV. 

A precise prediction for the $H\rightarrow ZZ $ partial width is obtained with Prophecy4f~\cite{Bredenstein:2006ha,Bredenstein:2006rh} (version 3.0~\cite{Denner:2019fcr}). This package includes the full QCD and EW NLO corrections to the Higgs boson decay width to four fermions, the interference contributions between different WW/ZZ channels, and all off-shell effects of intermediate W/Z bosons. The partial width $\Gamma(H\rightarrow ZZ)$ for our choice of parameters is 0.109~\MeV.  

The ratio of the $b\bar{b}$ and $ZZ$ partial widths obtained for $m_b(m_b) =$ 4.18~\gev{} is 21.76, in agreement within the uncertainty with the reference value of \mbox{22.0 $\pm$ 0.5} from Ref.~\cite{deFlorian:2016spz} used by ATLAS and CMS. The two results are fully compatible once the different input values for the Higgs boson mass are accounted for.

For our numerical analysis, the dependence of the ratio $\Gamma^{b\bar{b}}/\Gamma^{ZZ}$ on the bottom quark mass 
$m_b(m_H)$ is parameterized with a polynomial. 
The uncertainty in the fitted mass value due to the parameterization is below the per mille level over the mass range of interest. Variations of the functional form and fit range lead to negligible uncertainties.

\textbf{Impact of theory uncertainties.} The theory uncertainty on the bottom quark mass extraction from the Higgs coupling measurement due to missing higher orders is estimated following Ref.~\cite{deFlorian:2016spz} (and earlier work in Refs.~\cite{Heinemeyer:2013tqa,Denner:2011mq}). 
The uncertainties on the predicted ratio $\mu^{bb}/\mu^{ZZ}$ due to missing higher orders are estimated by varying the renormalization scale by factors between two and one half. Independent variations\footnote{The independent variations yield a slightly larger uncertainty compared to the 0.2\% reported in Ref.~\cite{deFlorian:2016spz}.} of the scales for $\alpha_{s}$ and $m_b$ yield a variation of $\sim$0.3\%. EW corrections beyond NLO are estimated to be $\sim$0.5\% on both partial widths \cite{deFlorian:2016spz}. 

The parametric uncertainty from $\alpha_s$ is estimated by propagating the 0.001 uncertainty on the PDG world average for $\alpha_s(m_Z)$ explicitly in HDECAY\footnote{The relatively small impact of the $\alpha_s$ uncertainty reported here can be understood. The larger effect reported in Ref.~\cite{deFlorian:2016spz} stems mainly from the evolution of the bottom quark mass from $\mu = m_b$ to $\mu = m_H$. This evolution is avoided altogether when $m_b(m_H)$ is adopted as the input parameter.}, which shifts the ratio of branching fractions by 0.2\%. 

The parametric uncertainty from the Higgs boson mass is estimated by varying the Higgs boson mass by $\pm$240~\mev{} around the nominal value $m_H=$ 125.1~\gev{} and recalculating the partial Higgs boson decay width to $Z$-boson pairs with Prophecy4f. Both the central value of the Higgs boson mass and the variations are based on the ATLAS + CMS Run 1 combination of Higgs boson mass measurements in the $\gamma\gamma$ and $ZZ$ decay channels~\cite{ATLAS:2015yey}. This leads to a variation of $\Gamma_{ZZ}$ by 3\% and is the dominant uncertainty on the ratio.

The linear sum of the several contributions yields a total theory uncertainty of 4.4\% on the $\Gamma^{bb}/\Gamma^{ZZ}$, which yields an uncertainty of 60~\MeV{} on $m_b(m_H)$. At the current experimental precision, the uncertainty on the theoretical prediction of the ratio is negligible.

\textbf{Extraction of $m_b(m_H)$ from Higgs rates.} We extract the bottom quark mass from the measurements of Eq.~\ref{eq:mubbmuzz_atlas} and Eq.~\ref{eq:mubbmuzz_cms}. The two results~\footnote{The intermediate result based on the ATLAS measurement~\cite{ATLAS-CONF-2020-027} is :
\begin{equation}
 m_b(m_H) =  2.61^{+0.32}_{-0.27} \mathrm{(stat.)} ^{+0.26}_{-0.19} \mathrm{(syst.)}\/  \gev,
\end{equation}
and that based on the CMS measurement~\cite{Sirunyan:2018koj} is:
\begin{equation}
m_b(m_H) = 2.57^{+0.39}_{-0.35} \mathrm{(stat.)} ^{+0.37}_{-0.28} \mathrm{(syst.)} \/  \gev.
\end{equation}}
are combined using the Convino package~\cite{Kieseler:2017kxl}, taking into account correlations among the asymmetric systematic uncertainties. The resulting value of the bottom quark mass is:
\begin{equation}
m_b(m_H) = 2.60^{+0.36}_{-0.31}~\gev.
\label{eq:mbmh}
\end{equation}
This is the main result of this Letter. 

\section{The running bottom quark mass}

The new measurements of $m_b(m_H)$ based on ATLAS and CMS measurements of Higgs decay rates (indicated with open red markers) and the average of both measurements ({\color{red} red $\star$}) are presented together with existing results for the bottom quark mass in Fig.\ref{fig:running_mass}. The PDG world average of $m_b(m_b)$ is indicated with a {\color{green} green $\star$} and the measurements of the bottom quark \MSbar{} mass at $m_Z$ by the LEP experiments and SLD with blue, open markers. 

\begin{figure}[th!]
\includegraphics[width=0.99\columnwidth]{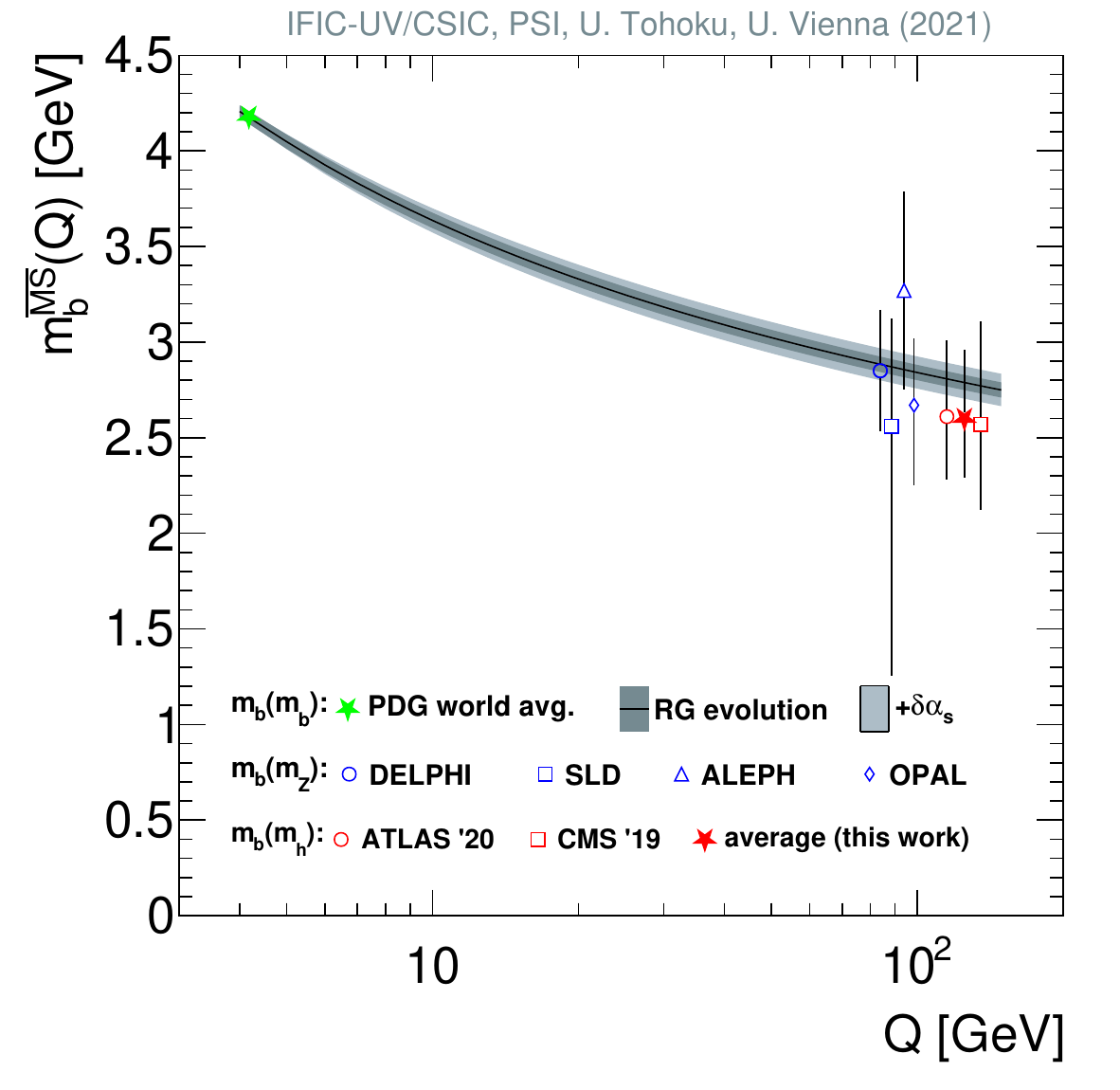}%
\caption{\label{fig:running_mass} The scale evolution of the bottom quark \MSbar{} mass. The measurements include the PDG world average for $m_{b}(m_{b})$ determined at a typical scale of the bottomonium mass, the measurements of $m_b(m_Z)$ from jet rates at the $Z$-pole at LEP and SLC and the measurement of $m_b(m_H)$ from Higgs boson branching fractions. The prediction of the evolution of the mass is calculated from the world average for $m_b(m_b)$ at five-loop precision with REvolver~\cite{Hoang:2021fhn}. The inner dark grey error band includes the effect of missing higher orders and the parametric uncertainties from $m_b(m_b)$ and $\alpha_s$ from the PDG averages. The outer band with a lighter shading includes additionally the effect of a $\pm$ 0.004 variation of $\alpha_s(m_Z)$. }
\end{figure}

The measurements at different scales are connected by the predicted RG evolution of $m_b(Q)$ in QCD. The evolution of the PDG world average for $m_b(m_b)$ to higher scales is given by the black curve, using the REvolver code~\cite{Hoang:2021fhn} at five-loop precision. The dark grey error band indicates the uncertainty on $m_b(Q)$ within the SM, with the dominant uncertainties stemming from the parametric $m_b(m_b)$ and $\alpha_s$ uncertainties~\cite{Zyla:2020zbs}. The impact of higher order uncertainties, estimated as half the difference between the three-loop and four-loop prediction, is negligible in comparison. 

The measurements at high scales are in good agreement with the evolution predicted by the SM. 

\textbf{The anomalous mass dimension.} The QCD scale evolution of the $\overline{\rm MS}$ quark masses $m_q(\mu)$ can be written in terms of the anomalous dimension $\gamma_m$ and the scale-dependent strong coupling $\alpha_s(\mu)$:
\begin{equation}
    \frac{\partial m_q(\mu)}{\partial \log({\mu^2})} = \gamma_m[\alpha_s(\mu)] \, m_q(\mu),
\end{equation}
Focusing on the first term in the expansion $\gamma_m[\alpha_s] = \gamma_0 \frac{\alpha_s}{\pi} +\mathcal{O}(\alpha^2_s)$, we obtain, in leading-log (LL) approximation:
\begin{equation}
    \gamma_0 = - \beta_0 \log\left({\frac{m_q(\mu^2)}{m_q(\mu^2_0)}}\right)/ \log\left({\frac{\alpha_s(\mu^2)}{\alpha_s(\mu^2_0)}}\right). 
\end{equation}
In the SM $\gamma_0 = -1$ and $\beta_0$ is given as a function of the (effective) number of flavours $n_f$. Adopting the five-flavour scheme, $\beta_0 = (33 - 2 n_f)/12 = 23/12$. The LL approximation is expected to be accurate to approximately 12\%, where the uncertainty is given by the size of the next-to-leading-log (NLL) correction within the five-flavour scheme.

Combining the values of $m_b(m_b)$, $m_b(m_Z)$ and $m_b(m_H)$ of Eq.~\ref{eq:mbmb},~\ref{eq:mbmz} and~\ref{eq:mbmh} one can determine $\gamma_0$ experimentally. We perform a $\chi^2$ minimization to obtain the following best-fit value for the ratio:
\begin{equation}
 \gamma_0/\beta_0 = -0.64 \pm 0.12 ({\rm exp.})  \pm 0.08 ({\rm theo.}) \pm 0.03 (\alpha_s).
 \end{equation}
The uncertainty due to $\alpha_s$ is evaluated by propagating the experimental uncertainties on $\alpha_s(m_b)$, $\alpha_s(m_Z)$ and $\alpha_s(m_H)$. To reduce the SM bias, a conservative uncertainty of 0.004 is assigned to the $\alpha_s$ values at $m_Z$ and $m_H$. This covers the envelope of experimental measurements of $\alpha_s$ at high scale from deep-inelastic scattering and parton distribution function fits as well as EW precision fits based on the pre-averaging quoted in Ref.~\cite{Zyla:2020zbs}.

With $\beta_0=23/12$, we find the following value for the anomalous mass dimension: 
 \begin{equation}
 \gamma_0 = -1.23 \pm 0.22 ({\rm exp.})\pm 0.14 ({\rm theo.})  \pm 0.06 (\alpha_s) ,
\end{equation}
in good agreement with the SM result. 

The leading-log approximation is found to be sufficiently accurate for the current measurement precision. A combined analysis of the evolution of the strong coupling and the bottom quark mass can disentangle the running of $\alpha_s$ and $m_b$ and may be an interesting diagnostic tool for new physics effects that impact their scale evolution in different ways.

\textbf{Testing the "running" hypothesis.} With the independent determinations of the bottom quark mass at different scales, we can test the hypothesis of the running of the bottom quark mass. To avoid a SM bias, we again relax the assumption that the evolution of the strong coupling is given by the RGE prediction, increasing the uncertainty on $\alpha_s$ at high scales to 0.004. The impact of this additional uncertainty is shown as a second, light grey band in Fig.~\ref{fig:running_mass}. Even with this additional uncertainty, the total uncertainty on the prediction for $m_b(m_H)$ is 90~\MeV, still more than three times smaller than the experimental uncertainty.

We test the running hypothesis with the following parameterization, adapted from Ref.~\cite{Sirunyan:2019jyn}:
\begin{multline}
  m(\mu;x,m_b(m_b)) =  \\
   x \Big[m^{\rm RGE}_b(\mu,m_b(m_b)) - m_b(m_b)\Big] + m_b(m_b),
  \label{eq:chisquared}
\end{multline}
where $m^{\rm RGE}_b(\mu,m_b(m_b))$ describes the RGE evolution expected in the SM for a reference mass $m_b(m_b)$, and $x$ is a multiplicative factor that adjusts the scale dependence, interpolating smoothly between the no-running scenario ($x=$ 0) and the SM ($x=$ 1).

We fit the predicted scale evolution of the bottom quark \MSbar mass $m$ with a $\chi^2$ minimization~\footnote{The $\chi^2$ is given by:
\begin{multline}
    \chi^2(x,m_b(m_b))= \\
    \sum_{\mu_i}(m(\mu_i;x,m_b(m_b))-m_b^{\rm exp}(\mu_i))^2/(\sigma_i)^2,
\end{multline}
where $m(\mu_i; x, m_b(m_b))$ is given by Eq.~\ref{eq:chisquared} and the experimental values $m_b^{exp}(\mu_i)$ are the averages for $m_b(m_b)$, $m_b(m_Z)$ and $m_b(m_H)$ given in Eq.~\ref{eq:mbmb}, Eq.~\ref{eq:mbmz} and Eq.~\ref{eq:mbmh}, respectively. The uncertainties $\sigma_i$ include the experimental uncertainties listed in Eq.~\ref{eq:mbmb}, Eq.~\ref{eq:mbmz} and Eq.~\ref{eq:mbmh}, and the parametric uncertainty due to $\alpha_s$ in the evolution.} using Eq.~\ref{eq:chisquared} and the averages of $m_b(m_b)$, $m_b(m_Z)$ and $m_b(m_H)$ given in Eq.~\ref{eq:mbmb}, Eq.~\ref{eq:mbmz} and Eq.~\ref{eq:mbmh}. The resulting $\chi^2$ values are shown in Fig.~\ref{fig:chisquared}, as a function of the two fit parameters. The best-fit value for the reference mass is $m_b(m_b) =$ 4.18$^{+0.03}_{-0.02}$ ~\gev{}, compatible with the PDG world average. The fit yields $x=$ 1.08 $\pm 0.15$ (exp.) $\pm 0.05$ ($\alpha_s$), where the first uncertainty corresponds to a propagation of the uncertainty on the mass measurements and the second to an uncertainty of $\pm0.004$ on the value of $\alpha_s(m_Z)$ used in the RG evolution. The best-fit value of $x$ is compatible with the SM prediction ($x=1$), within 1$\sigma$ and differs by nearly seven standard deviations from the no-running scenario ($x=0$). A fit of $m_b(m_b)$ and $m_b(m_Z)$, without the Higgs data, yields a value of $x$ of 1.03 $\pm$ 0.21, just below five standard deviations.

\begin{figure}[h!]
\includegraphics[width=0.99\columnwidth]{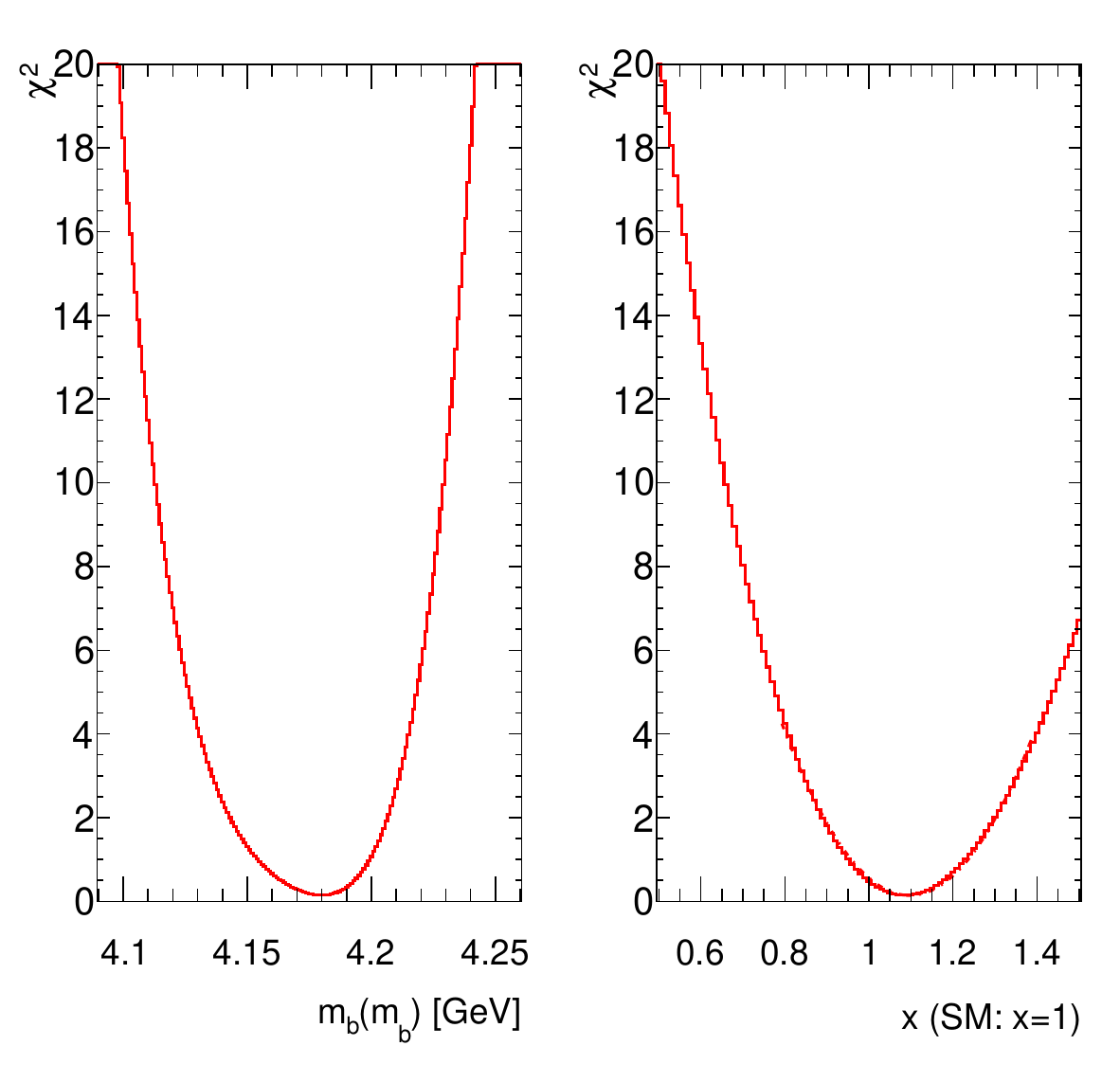}
\caption{\label{fig:chisquared} The $\chi^2$ of the fit of Eq.~\ref{eq:chisquared} to the measurements of $m_b(m_b)$, $m_b(m_Z)$ and $m_b(m_H)$, as a function of the reference bottom quark mass $m_b(m_b)$ and the factor $x$ that multiplies the RGE evolution to higher scale. The factor $x$ interpolates smoothly between the ``no-running'' scenario ($x=$ 0) and the RGE evolution predicted by QCD ($x=$ 1).}
\end{figure}

\section{Discussion \& outlook}

\textbf{Caveat.} When the Higgs decay rates are used for a determination of the bottom quark mass, we must assume that physics beyond the SM has a neligible impact. The procedure followed by the ATLAS and CMS experiments is quite robust against certain new physics effects. The contribution of unknown "invisible decays" to the Higgs width cancels in the ratio and other assumptions, e.g. on the Higgs boson production cross sections, can be tested to good precision. A shift of the bottom quark Yukawa coupling (and none of the other Higgs couplings) would, however, lead to a bias in the mass measurement. The results in this Letter are strictly valid only for a SM bottom quark Yukawa coupling.

\textbf{Prospects.} Future improvements of the Higgs branching fraction measurements are expected to rapidly reduce the uncertainties of this method. Projections for the HL-LHC~\cite{Cepeda:2019klc} envisage a measurement of $BR(H \rightarrow b\bar{b})$ with a precision of 4.4\%, reducing the experimental uncertainty on $m_b(m_H)$ to $\pm$60~\MeV.

The recoil mass analysis at a future electron-positron ``Higgs factory'' can reach sub-\% precision for Higgs boson couplings~\cite{deBlas:2019rxi,Barklow:2017suo,Abramowicz:2016zbo}, with minimal assumptions on the total width or the production rates. The ratio $\mu^{bb}/\mu^{WW}$ of the Higgs branching fractions (which is preferred over the $\mu^{bb}/\mu^{ZZ}$ ratio because of the larger branching fraction) is expected to be measured with 0.86\% precision for the 250~\gev{} stage of the International Linear Collider (ILC) and 0.47\% for the complete 250+500~\gev{} program~\cite{Fujii:2019zll,Bambade:2019fyw}, corresponding to an uncertainty on $m_b(m_H)$ of $\pm$12~\MeV~and $\pm$6~\MeV, respectively. Future $e^+e^-$ colliders furthermore offer opportunities to improve the precision of $m_b(m_Z)$, either with a dedicated high-luminosity run at the $Z$-pole or with radiative-return events, and to extend the analysis to $m_b(250~\gev)$~\cite{ILDnote2020}. 


\textbf{Summary.} In this Letter we have presented a new method to determine the bottom quark mass from the Higgs boson decay rate to bottom quarks and have used it to perform the first measurement of the bottom quark mass \MSbar{} mass at the renormalization scale of the Higgs boson mass. Combining ATLAS and CMS run 2 results, we obtain $m_b(m_H) = 2.60^{+0.36}_{-0.31}~\gev$,
in good agreement with the value $2.79^{+0.03}_{-0.02}~\gev$ expected from evolving the world average for $m_b(m_b)$ to the Higgs mass scale. The measurement combines a well-defined scale for the observable with a very small theory uncertainty and has excellent prospects at the HL-LHC and a future Higgs factory.

Combining our result with the world average for $m_b(m_b)$ and the determination of $m_b(m_Z)$ by the LEP and SLC experiments, we can test the "running" of the bottom quark mass. The observed scale evolution is compatible within errors with the RGE evolution predicted in QCD and the data strongly disfavours the no-running scenario.

\begin{acknowledgments}
\textbf{Acknowledgements:} The idea for this analysis goes back to LEP days and 
was revived in a DESY-IFIC workshop organized by K.~Lipka. We thank Junping Tian for his inputs on the ILC potential and Stefano Catani for valuable comments. The authors acknowledge support from projects FPA2015-65652-C4-3-R, FPA2017-84445-P and PGC2018-094856-B-100 (MICINN/FEDER), support from the U. Valencia and CSIC for H. Yamamoto, PROMETEO/2017/053, PROMETEO-2018/060 and CIDEGENT/2020/21 (Generalitat Valenciana), iLINK grant LINKB20065 (CSIC), and the FWF  Austrian  Science  Fund  Project  No.\ P28535-N27 and Doctoral Program No.\ W1252-N27. 
\end{acknowledgments}

\section{Supplementary material (1): previous bottom mass measurements and combination}

\textbf{World average of $m_b(m_b)$ measurements.} The most precise extractions of the bottom quark mass~\cite{Narison:2019tym,Peset:2018ria,Kiyo:2015ufa,Penin:2014zaa,Alberti:2014yda,Beneke:2014pta,Dehnadi:2015fra,Lucha:2013gta,Bodenstein:2011fv,Laschka:2011zr,Chetyrkin:2009fv} rely on the measurement of the mass of bottomonium bound states and the $e^+e^- \rightarrow $ hadrons cross section as experimental input, in combination with QCD sum rules and perturbative QCD calculations. Several lattice QCD groups have published results, the most recent of which reaches a precision of approximately 0.3\%~\cite{Bazavov:2018omf,Colquhoun:2014ica,Bernardoni:2013xba,Lee:2013mla,Dimopoulos:2011gx} (see also the FLAG report~\cite{Aoki:2019cca}). The "world average" for $m_b(m_b)$ of Eq.~\ref{eq:mbmb} provided by the Particle Data Group~\cite{Zyla:2020zbs} has a relative precision of less than 1\%. 

The typical energy scale of the measurements included in the world average is relatively low. For the precise measurements of Refs.~\cite{Narison:2019tym,Peset:2018ria,Kiyo:2015ufa,Penin:2014zaa,Alberti:2014yda,Beneke:2014pta,Dehnadi:2015fra,Lucha:2013gta,Bodenstein:2011fv,Laschka:2011zr,Chetyrkin:2009fv} the typical scale is the mass of the bottomonium bound states ($\sim$10~\gev). The average includes also inputs at low energy from HERA~\cite{H1:2018flt} and the Babar and Belle experiments at the B-factories~\cite{Schwanda:2008kw,Aubert:2009qda}. The results obtained at higher energies, that we discuss next, are not included in the world average for $m_b(m_b)$ of Eq.~\ref{eq:mbmb}. We can therefore safely assume that the experimental determinations at higher renormalization scales are independent. 



\textbf{Determination of $m_b(m_Z)$ at the Z-factories.} Bottom quark mass measurements at a much higher scale became possible at LEP and SLC, where jet rates and event shapes are sensitive to subleading mass effects. A practical method to extract the bottom-quark mass from $Z$-pole data was proposed in Ref.~\cite{Bilenky:1994ad}. Three independent groups completed the necessary next-to-leading order (NLO) theoretical calculation of the three-jet rate for massive quarks~\cite{Rodrigo:1997gy,Bilenky:1998nk,Rodrigo:1999qg,Bernreuther:1997jn,Brandenburg:1997pu,Nason:1997tz,Nason:1997nw} (an NNLO calculation for the three-jet rate in $e^+e^-$ collisions, without bottom-quark-mass effects, is available in Ref.~\cite{GehrmannDeRidder:2008ug}).

The first measurement of this type was performed by the DELPHI collaboration~\cite{Abreu:1997ey} using the LEP $Z$-pole data. Similar measurements were also performed with SLD~\cite{Brandenburg:1999nb,Abe:1998kr} data, and by ALEPH~\cite{Barate:2000ab}, OPAL~\cite{Abbiendi:2001tw} and DELPHI~\cite{Abdallah:2005cv,Abdallah:2008ac}. For our analysis we combine the most precise determinations from three-jet rates of each experiment, listed in Table~\ref{tab:mbmz}. The resulting average for $m_b(m_Z)$ is given by Eq.~\ref{eq:mbmz}.

\begin{table}[h!]
\caption{\label{tab:mbmz} Measurements of the bottom-quark \MSbar mass at the renormalization scale $\mu=m_{Z}$, from three-jet rates with bottom quarks in $e^+e^-$ collisions at the $Z$-pole at LEP and SLD. For ALEPH and DELPHI the hadronization uncertainty is added in quadrature with the experimental uncertainty to yield the total systematic uncertainty.}
\begin{tabular}{lcc}
experiment & $m_b({m_Z})$ [GeV]  \\ \hline
ALEPH\cite{Barate:2000ab}     & 3.27 $\pm$ 0.22 (stat.) $\pm$ 0.44 (syst.) $\pm$ 0.16 (theo.) \\
DELPHI\cite{Abdallah:2005cv}     & 2.85 $\pm$ 0.18 (stat.) $\pm$ 0.23 (syst.) $\pm$ 0.12 (theo.) \\  
OPAL\cite{Abbiendi:2001tw}    & 2.67 $\pm$ 0.03 (stat.) $^{+0.29}_{-0.37}$ (syst.) $\pm$ 0.19 (theo.)   \\ 
SLD\cite{Abe:1998kr,Brandenburg:1999nb}       &  2.56 $\pm$ 0.27 (stat.) $^{+0.28}_{-0.38}$ (syst.) $^{+0.49}_{-1.48}$ (theo.)                \\ \hline
\end{tabular}
\end{table}

The result is obtained with the Convino method~\cite{Kieseler:2017kxl}, propagating the reported asymmetric uncertainties and taking into account correlations between the systematic uncertainties of the measurements. Theory uncertainties are assumed to be 100\% correlated among the measurements, as all rely on the NLO prediction of the three-jet rates. The correlation among experimental systematic uncertainties is taken to be 50\%. This accounts for the correlated hadronization uncertainty, that is found to be of a similar magnitude as the statistical and purely experimental uncertainties in the ALEPH and DELPHI analyses that quote it separately. The combination is robust under variations of the assumed correlation: a combination performed with a 30\% or 70\% correlation among experimental systematic uncertainties yield a 20-40~\MeV\, variation of the average. The result obtained with the Best Linear Unbiased Estimator (BLUE~\cite{Nisius:2020jmf} version 2.4.0) and symmetrized uncertainties agrees with Eq.~\ref{eq:mbmz} within 20~\MeV.

\section{Supplementary material (2): Identification of the natural energy scale for Higgs decay}

\textbf{Independent scale variations.}

The renormalization scales of the strong coupling and the bottom quark mass can also be chosen independently. As long as both scales are of order $m_H$, the convergence of the series remains excellent and comparable to Eq.~\ref{Gammamumh}. On the other hand, using $\mu=m_H$ as the renormalization scale for the strong coupling and $\mu=m_b(m_b)=$ 4.18~\GeV{} as the renormalization scale for the bottom quark mass the leading perturbation series yields:
\begin{displaymath}
\label{GammamumbmH}
1 + \delta_{\rm QCD} = 1 - 0.2840 - 0.1178 - 0.0355  - 0.0079.
\end{displaymath}
The corresponding series, using the bottom pole mass (which can be considered to have an intrinsic renormalization scale of order $m_b$ in this context) with $m_b^{\rm pole}=4.9$~GeV and $\mu=m_H$ as the renormalization scale for the strong coupling, adopts the form:
\begin{displaymath}
\label{Gammambpole}
1 + \delta_{\rm QCD} = 1 - 0.3567 - 0.1521 - 0.0559  - 0.0239.
\end{displaymath}
Both series exhibit a better behavior than the approach of Eq.~\ref{Gammamumb}, but have substantially larger theoretical uncertainties than Eq.~\ref{Gammamumh}, indicating that both renormalization scales (for the bottom mass and the strong coupling) have to be chosen of the order of the Higgs mass to resum powers of $\ln(m_H/m_b)$ in an optimal manner.


Also the ${\cal O}(\alpha_s)$ corrections to the subleading $m_b^4/m_H^2$ contributions to the $H\rightarrow b\bar{b}$ partial width are free of powers of the large logarithm $\ln(m_H/m_b)$ and well-behaved only if a high renormalization scale such as $m_H$ is adopted for the bottom mass. This can be easily seen from their analytic expression which has the form $\propto m_b(\mu)^4/m_H^2[1 + \frac{20}{3}\frac{\alpha_s}{\pi}(1-\frac{6}{5}\ln(\frac{m_H}{\mu}))]$~\cite{Chetyrkin:1995pd}. Finally, we note that the absence of the large logarithm $\ln(m_H/\mu)$ in ths expression for $\mu=m_H$ also involves in a nontrivial way the kinematic bottom mass dependence of the Higgs decay phase space. This emphasizes that mass effects associated to phase space boundaries are not related to kinematic mass schemes (such as the pole mass which at ${\cal O}(\alpha_s)$ is related to the $\overline{\rm MS}$ mass via the relation $m_b^{\rm pole}=m_b(m_b)[1+ 4/3(\alpha_s/\pi)]$ in the high energy limit. This argument once again underlines that the bottom mass determined from the $H\rightarrow b\bar{b}$ partial width must be identified with $m_b(m_H)$.

\bibliography{main.bib}

\end{document}